\documentclass[reprint,prl,superscriptaddress]{revtex4-1}

\usepackage{amsmath,amsthm,amsfonts,amssymb,amscd,amsbsy}
\usepackage{graphicx}
\usepackage{array}
\usepackage{float}
\usepackage{enumerate}
\usepackage[nodayofweek]{datetime}
\usepackage[english]{babel}
\usepackage{mathtools}
\usepackage[toc,page]{appendix}
\usepackage{verbatim} %multi-line comments
\usepackage{amsthm} %theoreme & definitionen
\usepackage{enumitem}
\usepackage{enumerate}
\usepackage{bbm} %math symbols
\usepackage[normalem]{ulem} %dashed & pointed underline
\usepackage{bm}
\newcommand{\be}{\begin}
\newcommand{\e}{\end}
\newcommand{\beq}{\begin{equation}}
\newcommand{\eeq}{\end{equation}}

\renewcommand{\l}{\left}
\renewcommand{\r}{\right}
\newcommand{\set}[1]{\mathbb{#1}}
\newcommand{\curly}[1]{\mathcal{#1}}

\newcommand{\C}{\set{C}}
\newcommand{\Z}{\set{Z}}

\newcommand{\Lam}{\Lambda}
\newcommand{\gam}{\gamma}

\theoremstyle{definition}

\theoremstyle{remark}

\usepackage{tikz}

\usetikzlibrary{decorations.pathreplacing,decorations.markings}
\usepackage{xcolor}
\usepackage{xspace}

\tikzset{
  mid arrow/.style={postaction={decorate,decoration={
        markings,
        mark=at position .5 with {\arrow[#1]{stealth}}
      }}},
}

\newcommand{\hex}{\tikz[scale=0.18]{\draw[thick]
(0:1)--(60:1)--(120:1)--(180:1)--(-120:1)--(-60:1)--cycle;
}
}

\newcommand{\minihex}{\tikz[scale=0.1]{\draw[thick]
(0:1)--(60:1)--(120:1)--(180:1)--(-120:1)--(-60:1)--cycle;
}
}

\begin{document}
\title{Existence of a Spectral Gap in the Affleck-Kennedy-Lieb-Tasaki Model on the Hexagonal Lattice}

%\date{}

\author{Marius Lemm}
\email{mlemm@math.harvard.edu}
\affiliation{Department of Mathematics, Harvard University, 1 Oxford Street, Cambridge, Massachusetts 02138, USA}

\author{Anders W. Sandvik}
\email{sandvik@buphy.bu.edu}
\affiliation{Department of Physics, Boston University, 590 Commonwealth Avenue, Boston, Massachusetts 02215, USA}
\affiliation{Beijing National Laboratory for Condensed Matter Physics and Institute of Physics, Chinese Academy of Sciences, Beijing 100190, China}

\author{Ling Wang}
\email{lingwangqs@zju.edu.cn}
\affiliation{Zhejiang Institute of Modern Physics, Zhejiang University, Hangzhou 310027, China}

\begin{abstract}
The $S=1$ Affleck-Kennedy-Lieb-Tasaki (AKLT) quantum spin chain was the first rigorous example of an isotropic spin system in the Haldane phase. The conjecture that the $S=3/2$ AKLT model on the hexagonal lattice is also in a gapped phase has remained open, despite being a fundamental problem of ongoing relevance to condensed-matter physics and quantum information theory. Here we confirm this conjecture by demonstrating the size-independent lower bound $\Delta >0.006$ on the spectral 
gap of the hexagonal model with periodic boundary conditions in the thermodynamic limit. Our 
approach consists of two steps combining mathematical physics and high-precision computational physics. We first prove a mathematical finite-size criterion which 
gives an analytical, size-independent bound on the spectral gap if the gap of a particular cut-out subsystem of 36 spins exceeds a certain threshold 
value. Then we verify the finite-size criterion numerically by performing state-of-the-art DMRG calculations on the subsystem. 
\end{abstract}

\maketitle

The manifestations of antiferromagnetism in quantum spin systems depend sensitively on the underlying geometry and spin number. A subtle 
and famous instance of this connection was proposed by Haldane, who predicted in 1983 that the Heisenberg spin chain has a spectral gap above the ground state 
whenever the spin $S$ per site is an integer \cite{H83a,H83b}. Motivated by his considerations, Affleck, Kennedy, Lieb, and Tasaki (AKLT) introduced a 
new family of quantum spin systems in 1987 and proved that their one-dimensional $S=1$ version is indeed in Haldane's eponymous quantum phase \cite{AKLT87,AKLT88}. 
The influence of the seminal AKLT papers continues to this day: the valence-bond solid (VBS) aspect of the AKLT construction directly inspired the development 
of concepts that are by now central tenets of modern quantum physics, such as matrix product states, projected entangled pair states (PEPS), and more generally 
tensor network states \cite{Cetal1,Cetal2,FNW,O14,S11,Schuchetal1,Schuchetal2}. Moreover, the non-local string order exhibited by the AKLT 
chain \cite{KT,NR,Pollmannetal} has been developed much further into the more general concept of symmetry-protected topological order 
\cite{Chenetal,FK,Morimotoetal,Nussinov}. Finally, the AKLT ground states on some two-dimensional lattices, including the $S=3/2$ model on the 
hexagonal lattice, provide rare instances of a universal resource state for measurement-based quantum computation (MBQC) \cite{VC,WAR11,WHR14,M}. 

One of the main accomplishments of the original AKLT works \cite{AKLT87,AKLT88} is the rigorous derivation of a spectral gap above the AKLT ground state in one dimension. 
AKLT also investigated the $S=3/2$ model on the hexagonal lattice and were able to demonstrate the exponential decay of the spin-spin correlations for the exact VBS ground state with periodic boundary conditions, and on the basis of this fact they conjectured that the hexagonal model also exhibits a spectral gap (see also \cite{KLT88}). We recall that a spectral gap implies the decay of ground state correlations, but not vice-versa \cite{Fernandezetal1,Fernandezetal2,HK,N,NS}.
Evidence pointing to a spectral gap has been mounting \cite{Aetal,DB16,GMW,KLT88,K,LSY,PW19}, but, despite the paradigmatic role played by the
hexagonal AKLT model, the long-standing fundamental problem to show that its spectrum is gapped has remained unresolved. The presence of a gap would have broader consequences, 
e.g., in supporting the widespread heuristic that PEPS arise from gapped Hamiltonians, see the recent review \cite{Ciracetal}, and for the complexity and stability of the corresponding universal resource states for MBQC  \cite{VC,WAR11,WHR14,M}. One of the main reasons why the AKLT conjecture has remained unresolved is that, while the ground states of the hexagonal AKLT model can be written 
down exactly, only very little is known about its excited states. More generally, the existing mathematical techniques for deriving spectral gaps in quantum 
spin systems of dimensions $\geq 2$ are quite limited. The few examples where a spectral gap is known to exist include the product vacua with boundary states 
(PVBS) models \cite{BHNY,B,LN} and, since recently, decorated variants of the AKLT models \cite{Aetal,PW19}.

\begin{figure}[t]
\begin{center}
\includegraphics[width=65mm]{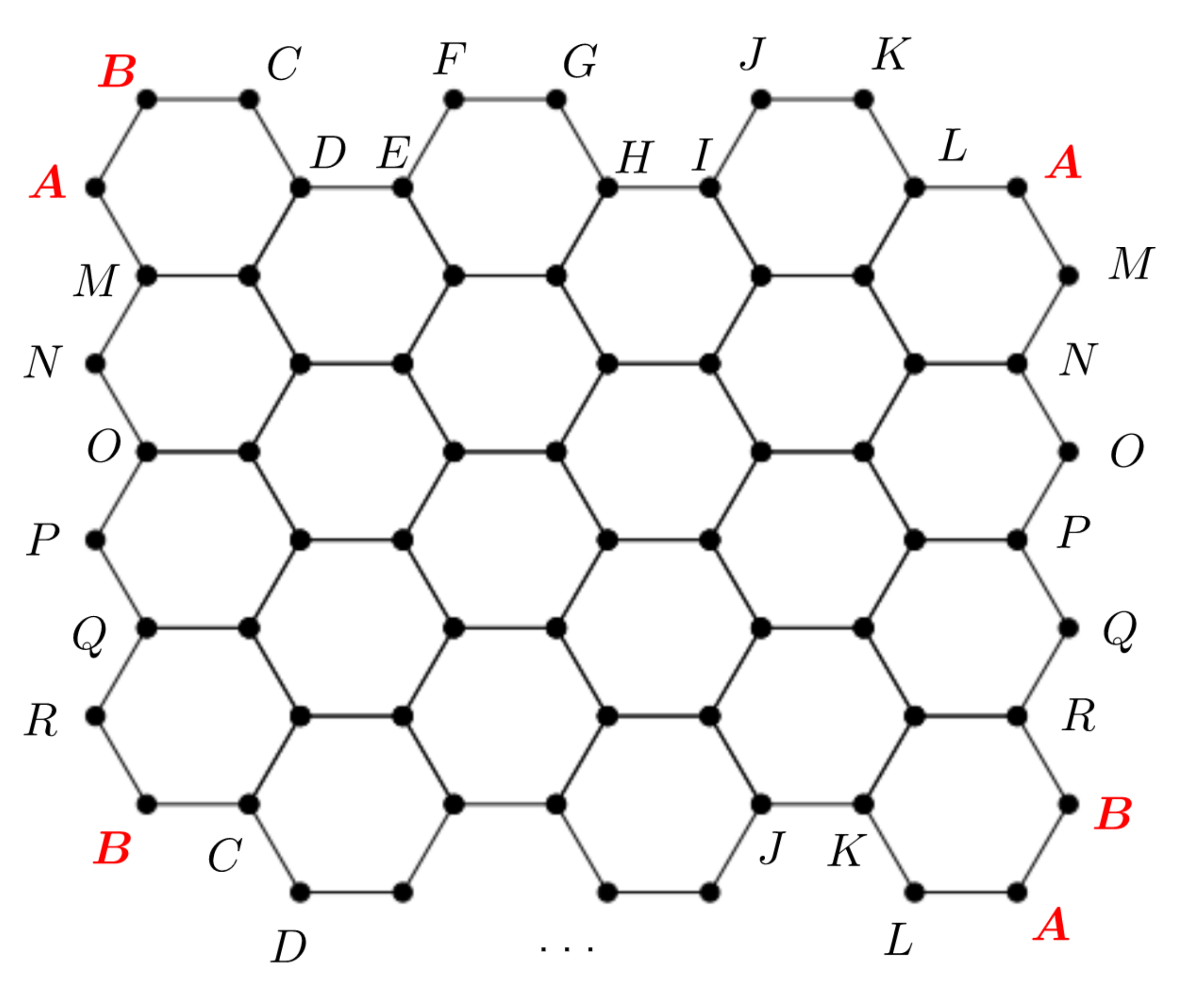}
\end{center}
\vskip-6mm
\caption{The patch $\Lam_{m_1,m_2}$ with parameters $m_1=6$ and $m_2=4$. ($m_1$ and $m_2$ are the width and height of $\Lam_{m_1,m_2}$ in units of hexagonal cells, respectively.) Periodic oundary conditions are imposed by identifying the boundary vertices which are assigned the same letter. Note that the letters
A and B appear three times in total.} 
%The resulting graph can equivalently be viewed as a hexagonal discretization of the torus.}
\label{fig:Lam}
\end{figure}

\begin{figure}[t]
\begin{center}
\includegraphics[width=52mm]{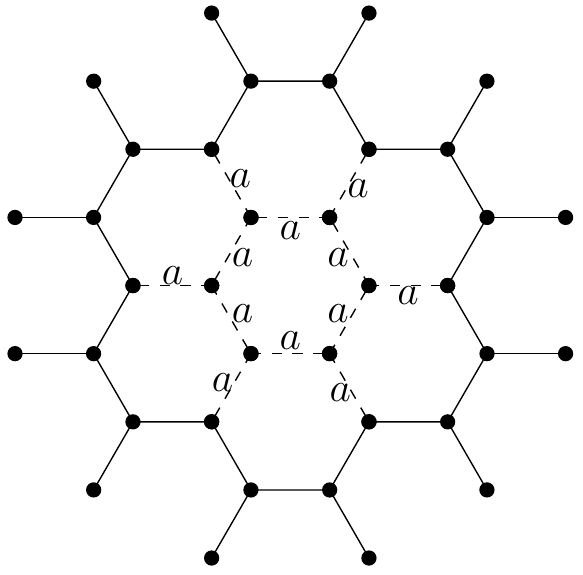}
\end{center}
\vskip-4mm
\caption{The fixed-size patch $\curly{F}$ whose spectral gap we compute numerically. It is equipped with open boundary conditions, in contrast to $\Lam_{m_1,m2}$. The weights $w_e$ in Eq.~\eqref{hakltedges} are assigned as follows: Dashed edges are weighted by $w_e=a\geq1$ as indicated, while all other edges 
are unweighted (i.e., $w_e=1$). }
\label{fig:F}
\end{figure}

In this Letter, we confirm the AKLT conjecture by demonstrating a lower bound, $\Delta > 0.006$, on the spectral gap of the hexagonal model. More precisely, we consider a sequence of AKLT models where the hexagonal lattice is wrapped on an $m_1\times m_2$ torus and show that their spectral 
gaps are all bounded from below by $0.006$ for arbitrarily large system-size parameters $m_1$ and $m_2$; see Fig.~\ref{fig:Lam} for the definition of the periodic boundary 
conditions on a $6\times 4$ torus. %As detailed below, the argument relies on numerical computations and therefore may not necessarily be considered a rigorous mathematical proof as a point of principle in the absence of rigorous error estimates.
Methodologically, our approach consists of two steps. Step 1 comes from mathematical physics and step 2 is based on 
state-of-the-art computational physics. In step 1, we prove a mathematical finite-size criterion which is tailor-made for the problem at hand. In a nutshell, 
the finite-size criterion says that, if the spectral gap of the $36$-site cluster displayed in Fig.~\ref{fig:F} exceeds an explicit 
numerical threshold, then the AKLT model has a spectral gap for all system sizes $m_1,m_2$. To prove the criterion, we follow the combinatorial approach 
pioneered by Knabe \cite{K}, strengthened by using interaction weights as in Refs.~\cite{GM,LM}. In step 2, we combine the rigorous analytical insight
from step 1 by numerically verifying the finite-size criterion via a high-precision density-matrix renormalization group (DMRG) calculation
(see also Ref.~\cite{LSY} for a one-dimensional analog studied with Lanczos diagonalization). We present tests of the correctness of our implementation of the
well-established DMRG method in the Supplemental Material (SM) [40]. Since
it is not possible to establish a rigorous precise estimate of any remaining
convergence errors, our result may not be
considered a rigorous mathematical proof as a matter of principle. However, in practice,
the computed gap exceeds the threshold by such a wide
margin that it can be regarded as a conclusive demonstration.

One challenge in the numerical part of the argument is that the relevant open-boundary system (Fig.~\ref{fig:F}), whose gap we need to compute, has a massive 
ground state degeneracy due to the  12 ``dangling'' effective boundary $S=1$ spins which arise in the AKLT construction when only one out of the three nearest-neighbor couplings is active. This results in a $3^{12}$-fold ground state degeneracy. To reduce the number 
of levels which has to be converged, we use a variant of DMRG with full $\mathrm{SU}(2)$ symmetry and calculate the ground state and several excited states over all 
sectors of total spin. Crucially, in the process of successively orthogonalizing the calculations to previously converged states, we have used the AKLT
construction to exactly project out the full degenerate subspace. Without this preliminary step, which we discuss further below and describe in more 
detail in \cite{SM}, it would currently not be possible to converge the excited states in all total spin ($J$) sectors 
and conclusively identify the smallest gap of the system. We find that the lowest gap originates from the $J=13$ sector and that it exceeds the analytical gap threshold well beyond any conceivable remaining DMRG truncation errors. %To avoid any misconceptions, we nonetheless wish to state clearly that the main result relies on this numerical data and, in the absence of rigorous error bounds, may not necessarily be considered a rigorous mathematical proof depending on one's standards of rigor.

Our main result is a size-independent lower bound on the spectral gap of the AKLT Hamiltonian on finite patches of the hexagonal lattice $\mathbb H$ with periodic boundary conditions, which we call $\Lam_{m_1,m_2}$. The key point is that the lower bound on the gap is independent of the size parameters $m_1$ and $m_2$ of these patches and thus extends to the thermodynamic limit.

For $m_1$ and $m_2$ two positive integers, the finite patch $\Lam_{m_1,m_2}$ is defined by wrapping the hexagonal lattice on an $m_1 \times m_2$ torus. 
We invite the reader to view Fig.~\ref{fig:Lam} for a specific example of how the periodic boundary conditions are realized. 
Since the hexagonal lattice has valence $3$, one takes each site to host an $S=3/2$ spin and considers the Hilbert space
\beq\label{eq:Hilbertdefn}
\curly{H}_{m_1,m_2}=\bigotimes_{j\in\Lam_{m_1,m_2}} \C^{4}.
\eeq
On $\curly{H}_{m_1,m_2}$, the AKLT Hamiltonian is defined by
\beq
H^{AKLT}_{m_1,m_2}=\sum_{\substack{j,k\in \Lam_{m_1,m_2}:\\ j\sim k}} P_{j,k}^{(3)},
\eeq
where $P_{j,k}^{(3)}$ denotes the projection onto total spin $3$ across the bond connecting vertices $j$ and $k$. By convention, the neighboring relation $\sim$ includes the periodic boundary conditions inherent to $\Lam_{m_1,m_2}$. 

As a sum of projections, the Hamiltonian $H^{AKLT}_{m_1,m_2}$ is automatically a positive semidefinite operator. The valence-bond construction of AKLT \cite{AKLT87,AKLT88} yields a ground state which is a non-zero element of $\ker H^{AKLT}_{m_1,m_2}$, making this Hamiltonian frustration-free. Its spectral gap 
$\gam^{AKLT}_{m_1,m_2}$ is the smallest strictly positive eigenvalue, that is,
\beq
\gam^{AKLT}_{m_1,m_2}=\inf\mathrm{spec} \l(H^{AKLT}_{m_1,m_2}\r)\setminus\{0\}.
\eeq

We can now state our main result, which provides a lower bound on the spectral gap $\gam^{AKLT}_{m_1,m_2}$ that is independent of the system size parameters $m_1$ and $m_2$.
\vspace{7pt}

\textbf{Main result.}
\textit{Let $m_1,m_2\geq 12$. Then, it holds that}
\beq\label{eq:main}
\gam^{AKLT}_{m_1,m_2}\geq 0.00646.
\eeq

\noindent
A few remarks about this result are in order: (i)  We work with periodic boundary conditions for convenience and the results imply a bulk gap in
the thermodynamic limit under these boundary conditions. Moreover, it was proved in Ref.~\cite{KLT88} that the infinite-volume ground state is unique.  
(ii) This main result is not a rigorous mathematical theorem because it relies on numerical input from the DMRG algorithm. While the DMRG algorithm becomes exact for large bond dimension and the computations are sufficiently precise and well-tested to firmly establish \eqref{eq:main} beyond doubt, we do not claim to have a mathematical proof of sufficiently tight error estimates. %Looking forward, it would be interesting to have a purely analytical proof of a spectral gap for this model, especially considering that such a proof would presumably provide structural information on the low-lying excited states.
(iii) From previous numerical investigations, see e.g.~\cite{GMW}, it is believed that the true spectral gap 
of the hexagonal model is $\approx 0.1$, but the results depend on extrapolations in the system size that assume that an asymptotic 
scaling regime has been reached.

{\it The finite-size criterion.}---We now discuss the main mathematical tool, which is a finite-size criterion for deriving a spectral gap. In a nutshell, it says that if the spectral gap of the system $\curly{F}$ depicted in Fig.~\ref{fig:F} exceeds some explicit numerical threshold, then we also obtain a lower bound on the spectral gap $\gam_{m_1,m_2}^{AKLT}$ that is independent of the size parameters $m_1,m_2$ as desired. The intuition behind the finite-size criterion is that, thanks to the frustration-freeness of the AKLT Hamiltonian, the problem of finding the lowest possible excitation energy (gap) is a local question. Hence, it is enough to know that local patches of the whole system are ``sufficiently gapped'' in a way that the criterion makes precise. For related finite-size criteria that ours here is inspired by, see Refs.~\cite{K,GM,LM,L1,Anshu,L2,LSY}. The idea behind the finite-size criterion is to construct $H_{m_1,m_2}^{AKLT}$ from translated copies of an appropriate finite-size Hamiltonian, which we call $H_{\curly{F}}$. For the criterion to work in practice, the patch has to be sufficiently large because the criterion depends on the cluster size and shape, and even 
if there is a gap in the thermodynamic limit  the finite-size criterion may not be satisfied on a small cluster. Our criterion is based on the following Hamiltonian $H_{\curly{F}}$ 
defined on the 36-site patch $\curly{F}$ shown in Fig.~\ref{fig:F}, with open boundary conditions. %Some of the local interactions comprising 
%$H_{\curly{F}}$ are weighted by a factor $a\geq 1$ (which we will set to $a=1.4$ eventually). 

The patch lives on the local Hilbert space $\curly{H}_{\curly{F}}=\bigotimes_{j\in \curly{F}} \C^4.$
We write $\curly{E}_{\curly{F}}$ for the set of edges $e=(j,k)$ with $j,k\in \curly{F}$, i.e., we equip $\curly{F}$ with open boundary conditions (in contrast to $\Lam_{m_1,m_2}$). Let $a\geq 1$ be a parameter. We define the finite-size Hamiltonian by
\begin{equation}
H_{\curly{F}}=\sum_{e\in \curly{E}_{\curly{F}}} w_{e} P_{e}^{(3)},
\label{hakltedges}
\end{equation}
where $P_{e}^{(3)}$ is the projection onto total spin $3$ for the pair of vertices $j,k$ that form the endpoints of the edge $e$.
The weights $w_{e}$ are defined as follows:
\beq\label{eq:wedefn}
w_{e}=
\be{cases}
a, &\textnormal{if the edge $e$ is labeled by $a$ in Fig.~\ref{fig:F}},\\
1, &\textnormal{otherwise}.
\e{cases}
\eeq

The valence-bond ground state construction of AKLT \cite{AKLT87,AKLT88} still applies to $H_{\curly{F}}$ and proves that it is frustration-free. Its spectral gap is $\gam_{\curly{F}}(a)=\inf\mathrm{spec} \l(H_{\curly{F}}\r)\setminus\{0\}.$

\be{thmnon}[The finite-size criterion]
Let $m_1,m_2\geq 12$ be integers and let $a\geq 1$. Then we have the gap bound
\beq\label{eq:fs}
\gam^{AKLT}_{m_1,m_2}\geq \frac{10+4a}{3a^2+2a+7}\l(\gam_{\curly{F}}(a)-\frac{a^2-2a+3}{10+4a}\r).
\eeq
\e{thmnon}

The general way of applying this theorem goes as follows: If for some parameter value $a\geq 1$, one finds that the finite--size gap $\gam_{\curly{F}}(a)$ exceeds the threshold $\frac{a^2-2a+3}{10+4a}$, then \eqref{eq:fs} provides a lower bound on $\gam^{AKLT}_{m_1,m_2}$ that is independent of $m_1,m_2$ (subject to $m_1,m_2\geq 12$ of course). The proof of the finite-size criterion is deferred to the SM \cite{SM}. 

We now follow this procedure to show the spectral gap bound \eqref{eq:main}. As explained in detail further below, by a numerical DMRG calculation we obtain 
the following explicit lower bound on the finite-size gap $\gam_{\curly{F}}(a)$ with $a=1.4$,
\beq\label{eq:gapvalue}
\gam_{\curly{F}}(1.4) > 0.145.
\eeq
This value exceeds the gap threshold $\frac{a^2-2a+3}{10+4a} \approx 0.138$, and thus verifies the 
finite-size criterion. The exact numerical bound on $\gam^{AKLT}_{m_1,m_2}$ can be computed by noting that
$
\frac{a^2-2a+3}{10+4a}<0.1385$ and $\frac{10+4a}{3a^2+2a+7}>0.994$,
which together with  \eqref{eq:gapvalue} can be applied to \eqref{eq:fs} to show
$$
\gam^{AKLT}_{m_1,m_2}\geq 0.994 \l(0.145-\frac{a^2-2a+3}{10+4a}\r)\geq 0.00646
$$
This establishes the main result, the spectral gap bound \eqref{eq:main}.

{\it DMRG calculations.}---We next discuss our implementation of the DMRG
algorithm and results for the gap of the open boundary 36-site cluster $\curly{F}$ shown in Fig.~\ref{fig:F}. Additional details, including 
detailed convergence tests, are relegated to the SM \cite{SM}. 

The ground states of the cluster $\curly{F}$ can be understood as follows: each physical
$S=3/2$ spin is made out of 3 auxiliary $S=1/2$ spins, each of which will pair with
another auxiliary $S=1/2$ from a neighboring site, forming a singlet and dropping
out. This construction ensures that any pair of neighboring physical $S=3/2$ 
spins can never fuse into a total spin-$3$ state, and the AKLT ground state condition
is therefore fullfilled. However on the open boundary sites, two auxiliary 
$S=1/2$ spins per site are left over, and these are only allowed to fuse into an 
$S=1$ state due to the symmetric constraint. Therefore, there are 12 boundary 
$S=1$ degrees of freedom that can form any total spin $0\le J\le 12$, 
spanning a degenerate ground state manifold of dimension $3^{12}$. The lowest 
excitation above the ground states, which can be interpreted as swapping a bulk singlet 
with a triplet that further fuses with the boundary total angular momentum, 
can in principle form any angular momentum $0\le J\le 13$. In order to conclusively determine the smallest 
nonzero gap among all possible total-spin sectors, one has to find the lowest excitation 
in every sector $J\in \{ 0,1,\ldots,13\}$. For even higher $J$ sectors, the lowest excitation requires 
breaking more than one singlet and therefore costs significantly more energy. For completeness
we also computed the gaps in all other sectors where $J>13$.

\begin{figure}[t]
\begin{center}
\includegraphics[width=65mm]{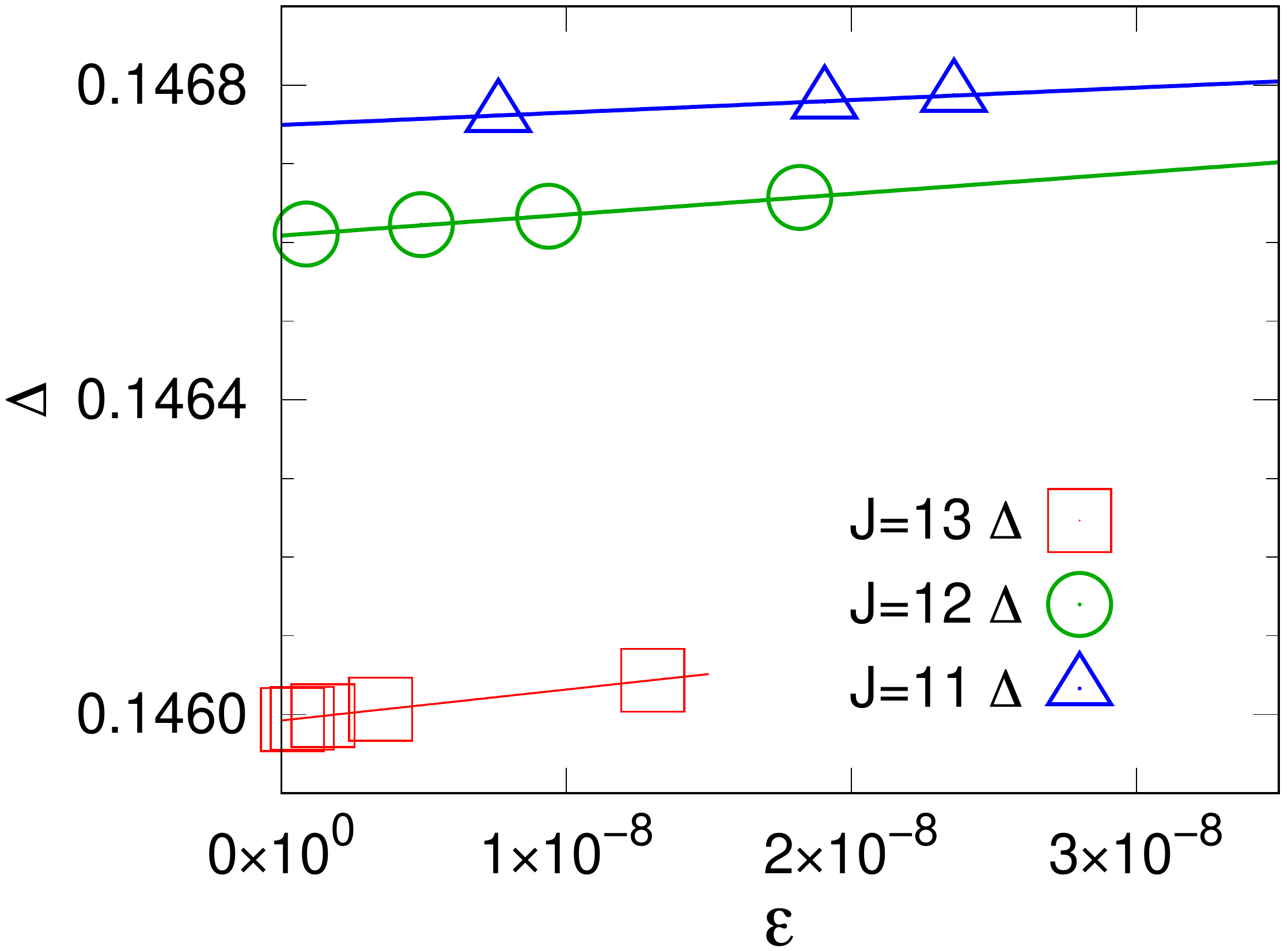}
\end{center}
\vskip-4mm
\caption{Gaps in the sectors $J=11,12$, and $13$ graphed versus the DMRG discarded weight $\epsilon$.
The discarded weight decreases with increasing number of $\mathrm{SU}(2)$ states used, and we used up to $D=2400$ for $J=11,12$, and up to $D=1200$ for $J=13$. Line fits are used for $\epsilon \to 0$ extrapolation.}
\label{dmrgresults}
\end{figure}

An $\mathrm{SU}(2)$ symmetric DMRG algorithm is used to automatically generate the
degenerate ground state manifold in all sectors of total spin
$J\in \{0,1,\ldots, 12\}$ and compute the lowest excited state therein by projecting out
the complete ground state manifold exactly. Two of us previously used such
an orthogonalization procedure for successively converging excited states of
a different model \cite{prl121.107202}, but here the simple form of the degenerate 
AKLT ground-state manifold enables us to eliminate it directly. Let $L$ denote the maximum-spin multiplet formed by the unpaired boundary $S=1$ 
spins in the ground state manifold. For the 36-site cluster in Fig.~\ref{fig:F}
we have $L=12$. The ground state manifold contains the following number of states
with total spin $J$: 4213 ($J=0$), 11298 ($J=1$), 15026 ($J=2$), 14938
($J=3$), 12078 ($J=4$), 8162 ($J=5$), 4642 ($J=6$), 2211 ($J=7$), 869
($J=8$), 274 ($J=9$), 66 ($J=10$), 11 ($J=11$), and 1 ($J=12$). Accordingly, the lowest
excitation for each $J$ is computed by projecting out that many degenerate 
ground states, which make the excited state computationally challenging. For sectors with total spin $J>L$,
which are devoid of ground states, the lowest excitation can be computed more straight-forwardly
without projecting out any states. Upon computing the lowest excitation gaps
for all $J\le L+1$ sectors of the 36-site cluster at $a=1.4$, we found that the smallest 
one originates from the $J=L+1=13$ sector; in Fig.~\ref{dmrgresults} we show results
for $J=11,12$, and $13$. The $J=13$ gap obtained by extrapolating to vanishing DMRG 
discarded weight $\epsilon$ is $\Delta(13)=0.14599$. The lowest gaps within all other $J$ 
sectors remain well above $\Delta(13)$ and there is no doubt (but also no rigorous proof) that the smallest gap exceeds 
the relevant threshold $0.138$. In the SM, the convergence of the gaps with $\epsilon$ is 
illustrated in Fig.~\ref{fig:dmrg} for all $0\le J\le 16$.

{\it Conclusions.}---We have verified the AKLT conjecture from 1987 that the hexagonal AKLT model has 
a spectral gap  above the ground state. This confirms that the original Hamiltonian with a PEPS ground state is gapped, a question emphasized, e.g., in the recent collection of open problems \cite{Ciracetal}. More generally, the existence of a spectral gap is an immensely consequential
property in any quantum many-body system. First, a spectral gap implies the exponential decay of ground state correlations (but not vice-versa) \cite{Fernandezetal1,Fernandezetal2,HK,N,NS} and is expected to imply other complexity bounds on the ground state such as the area law for the entanglement entropy. Second, the existence of a spectral gap is a crucial assumption in the classification of topological quantum phases and the many-body adiabatic theorem \cite{BBDF,BDRF,BMNS12,BHM10,H04,HW}.  We also mention that the existence of a spectral gap is perturbatively stable \cite{BHM10,H19,dRS,FP,MZ}. %In summary, the existence of a spectral gap shown here will play a key role in making precise the quantum phase exhibited by the AKLT ground state on the hexagonal lattice. 
While our result confirms the long-standing AKLT conjecture, we hope that it inspires future 
work on the spectral gap of this timeless model. In particular, we believe that it would be useful 
to have a purely analytical derivation of a spectral gap, because the argument here relies on numerical computations without suitable rigorous error bounds and because a purely analytical argument will presumably be accompanied by an improved understanding of the model's low-energy excitations. 

Let us briefly discuss the wider scope of the approach we use here. The mathematical physics step is the derivation of a finite-size criterion in the general spirit of Knabe's combinatorial criteria \cite{K} with weights as in Refs.~\cite{GM,LM}. The computational physics step consists of verifying the finite-size criterion by a high-precision DMRG implementation. Our approach of numerically verifying a combinatorial finite-size criterion is in principle applicable to any frustration-free spin system. Concerning the AKLT models, for example, the square lattice is a natural next candidate to consider \cite{AAH,GMW,PW19}, as well as $\mathrm{SU}(n)$-symmetric variants \cite{GS,WNM,GP}. The cubic lattice is another interesting case which also displays novel phase-transition phenomena \cite{PSA09}. %At any rate, the design and verification of the finite-size criterion is performed on a case-by-case basis. 

\textit{Note added: } After our preprint appeared, Pomata and Wei \cite{PWrecent} demonstrated the existence of a spectral gap in AKLT models on various two-dimensional degree-$3$ lattices including the hexagonal lattice. Their argument is different, but it also combines analytics (inspired by \cite{Aetal,PW19}) with numerics. 

%Concerning the design of a useful (verifiable) finite-size criterion, we make the following remark. The gap threshold 
%[the fraction $\frac{a^2-2a+3}{10+4a}$ in \eqref{eq:fs} in our case] decreases as the finite-size system increases, because 
%it quantifies how efficiently a large system is covered by the translates of the finite-size system. The choice of finite-size 
%system thus needs to negotiate two competing requirements: (i) The finite-size system should be large enough, so that the gap 
%threshold is reasonably small. (ii) The finite-size system should be small enough, so that its spectral gap can be explicitly 
%computed using some reliable numerical technique. Of course, the design problem is in fact more subtle and there exist additional 
%principles, such as avoiding finite-size systems which may host edge modes of low-energy that spoil a bulk gap, that may differ 
%from system to system.

\section*{Acknowledgments}
We would like to thank Daniel Arovas for useful discussions.
ML thanks Bruno Nachtergaele for encouragement and advice. AWS was supported
by the NSF under Grant No.~DMR-1710170 and by a Simons Investigator Grant. LW was
supported by the the National Natural Science Foundation of China, Grants
No.~NSFC-11874080 and No.~NSFC-11734002.

\begin{widetext}

\clearpage
  
\begin{center}  

\section{Supplementary Material}

{\bf\large Existence of a spectral gap in the AKLT model on the hexagonal lattice}
\vskip5mm

Marius Lemm, Anders W. Sandvik, and Ling Wang
\vskip3mm

\end{center}

In Section I, we present the detailed proof of the finite-size criterion. In Section II, we explain details concerning the 
implementation of the $\mathrm{SU}(2)$ symmetric DMRG method. In Section III, we demonstrate the correctness of our method of exactly projecting out the
ground state manifold on a 12-site cluster, for which the gaps can be computed exactly. We also demonstrate the convergence of the lowest gaps in 
all total-spin sectors $J \in \{0,1,\ldots 54\}$ of the  36-site cluster on the basis of which our conclusions on the numerical 
gap bound is drawn.
\vskip4mm
\end{widetext}

\setcounter{page}{1}
\setcounter{equation}{0}
\setcounter{figure}{0}
\setcounter{table}{0}
\renewcommand{\theequation}{S\arabic{equation}}
\renewcommand{\thetable}{S\Roman{table}}
\renewcommand{\thefigure}{S\arabic{figure}}

\subsection{I. Proof of the finite-size criterion}

\subsection{Squaring the Hamiltonian}

\noindent
Fix two integers $m_1,m_2\geq 12$. In the following, we abbreviate $H^{AKLT}_{m_1,m_2}=H$, $P^{(3)}_e=P_e$ and $\gam_{\curly{F}}(a)=\gam_{\curly{F}}$. 

By frustration-freeness and the spectral theorem, the claimed gap inequality \eqref{eq:fs} is equivalent to the operator inequality
\beq\label{eq:H2claim}
H^2\geq \frac{10+4a}{3a^2+2a+7}\l(\gam_{\curly{F}}-\frac{a^2-2a+3}{10+4a}\r)  H.
\eeq
As usual, an operator inequality $A\leq B$ is defined to mean that the operator $B-A$ is positive semidefinite.

Our goal is now to prove Eq.~\eqref{eq:H2claim}. We begin by computing $H^2$. Let us introduce some convenient notation. We write $\curly{E}_{m_1,m_2}$ for the set of edges of $\Lam_{m_1,m_2}$ considered \textit{with} periodic boundary conditions. Given two distinct edges $e$ and $e'$, we write $e\sim e'$ if $e\neq e'$ and the edges share a vertex, and we write $e\not\sim e'$, if $e\neq e'$ and the edges do not share a vertex. We also introduce the notation
$$
\{A,B\}=AB+BA
$$
for the anticommutator of two operators $A$ and $B$. 

Using that $P_e^2=P_e$, we have
\beq\label{eq:H2}
H^2=H+Q+R,
\eeq
where we introduced the operators
\beq\label{eq:QRdefn}
\begin{aligned}
Q=&\sum_{\substack{e,e'\in \curly{E}_{m_1,m_2}:\\ e\sim e'}} \{P_e,P_{e'}\},
\\
 R=&\sum_{\substack{e,e'\in \curly{E}_{m_1,m_2}:\\ e\not \sim e'}} \{P_e,P_{e'}\}.
\end{aligned}
\eeq
%For later, we observe that all terms $\{h_e,h_{e'}\}$ contributing to $R$ are non-negative, since the corresponding projections commute.

\subsection{Shifted finite-size systems and the auxiliary operator}
The idea is to construct the full Hamiltonian $H$ from translated copies of the finite-size Hamiltonian $H_{\curly{F}}$, viewed as subsystems acting on the common Hilbert space $\curly{H}_{m_1,m_2}$ from \eqref{eq:Hilbertdefn}. 

Let us introduce some formal setup and notation. We write $\curly{P}_{m_1,m_2}$ for the set of plaquettes in $\Lam_{m_1,m_2}$. Given a fixed plaquette $\hex\in \curly{P}_{m_1,m_2} $, we write $\curly{F}_{\minihex}$ for a copy of the patch $\curly{F}$ which has $\hex$ as its central plaquette and otherwise respects the periodic boundary conditions imposed by $\Lam_{m_1,m_2}$. 
%See Figure \ref{fig:boundary} for an example where $\curly{F}_{\minihex}$ intersects the boundary of $\Lam_{m_1,m_2}$. 
The edge set $\curly{E}_{\curly{F}_{\minihex}}$ is then defined accordingly, i.e., it respects the periodic boundary conditions of $\Lam_{m_1,m_2}$ and also the open boundary conditions of $\curly{F}_{\minihex}$. (Here we use that $m_1,m_2\geq 12$, so that these boundary requirements do not interfere.)

On the common Hilbert space $\curly{H}_{m_1,m_2}$ from \eqref{eq:Hilbertdefn}, we can then define the family of translated finite-size Hamiltonians
$$
H_{\curly{F}_{\minihex}}=\sum_{e\in \curly{E}_{\curly{F}_{\minihex}}} w_{e} P_{e}, \qquad \textnormal{ for every } \hex\in \curly{P}_{m_1,m_2}.
$$
We observe that these Hamiltonians are all unitarily equivalent to $H_{\curly{F}}$. In particular, they are frustration-free and their spectral gaps are all equal to $\gam_{\curly{F}}.$

We introduce the auxiliary operator
$$
\curly{A}=\sum_{\minihex\in \curly{P}_{m_1,m_2}} H_{\curly{F}_{\minihex}}^2.
$$

We have the following key lemma.

\be{lm}\label{lm:key}
Let $m_1,m_2\geq 12$ be integers and let $a\geq1$. 

We have the two operator inequalities
\begin{align}
\label{eq:A1}
\curly{A}\geq& (10+4a) \gam_{\curly{F}} H,\\
\label{eq:A2}
\curly{A}\leq& (10+4a^2) H+(3a^2+2a+7) (Q+R).
\end{align}
\e{lm}

\be{proof}
We first prove \eqref{eq:A1}. By frustration-freeness and the spectral theorem, it holds that
$$
H_{\curly{F}_{\minihex}}^2
\geq \gam_{\curly{F}_{\minihex}} H_{\curly{F}_{\minihex}}=\gam_{\curly{F}} H_{\curly{F}_{\minihex}}.
$$
In the second step, we used that $\gam_{\curly{F}_{\minihex}}=\gam_{\curly{F}}$ by unitary equivalence of the corresponding Hamiltonians. When we sum this operator inequality over plaquettes $\hex\in \curly{P}_{m_1,m_2}$, we find
\beq\label{eq:ALB}
\curly{A}\geq \gam_{\curly{F}} \sum_{\minihex\in \curly{P}_{m_1,m_2}} H_{\curly{F}_{\minihex}}
=\gam_{\curly{F}}\sum_{\minihex\in \curly{P}_{m_1,m_2}} \sum_{e\in \curly{E}_{\curly{F}_{\minihex}}} w_{e} P_{e}.
\eeq
By translation invariance, each $e\in \curly{E}_{m_1,m_2}$ appears the same number of times in the combined summation $\sum_{\minihex\in \curly{P}_{m_1,m_2}} \sum_{e\in \curly{E}_{\curly{F}_{\minihex}}}$, where we also account for the number of times the edge is accompanied by the weight factor $a$ arising from \eqref{eq:wedefn}. In other words, the sum of local Hamiltonians $\sum_{\minihex\in \curly{P}_{m_1,m_2}} H_{\curly{F}_{\minihex}}$ is a multiple of the full Hamiltonian $H$, where the multiplicative factor reflects the weighted number of times each edge appears in a copy of $\curly{F}_{\minihex}$. We find that a given edge $e\in \curly{E}_{m_1,m_2}$ appears $10$ times as an unweighted edge in a $\curly{F}_{\minihex}$, and $4$ times as an $a$-weighted edge. These combinatorial considerations show that
\beq\label{eq:sum}
\sum_{\minihex\in \curly{P}_{m_1,m_2}} H_{\curly{F}_{\minihex}}=(10+4a) H,
\eeq
which together with \eqref{eq:ALB} proves \eqref{eq:A1}.\\

It remains to prove \eqref{eq:A2}. Since $P_e^2=P_e$, we have as in \eqref{eq:H2},
$$
H_{\curly{F}_{\minihex}}^2=\tilde H_{\curly{F}_{\minihex}}+Q_{\curly{F}_{\minihex}}+R_{\curly{F}_{\minihex}},
$$
with
\beq\label{eq:QRminihexdefn}
\begin{aligned}
\tilde H_{\curly{F}_{\minihex}}=&\sum_{\substack{e,e'\in \curly{E}_{\curly{F}_{\minihex}}:\\ e\sim e'}}  w_e^2 P_e,\\
Q_{\curly{F}_{\minihex}}=&\sum_{\substack{e,e'\in \curly{E}_{\curly{F}_{\minihex}}:\\ e\sim e'}}  w_e w_{e'}\{P_e,P_{e'}\},\\
R_{\curly{F}_{\minihex}}=&\sum_{\substack{e,e'\in \curly{E}_{\curly{F}_{\minihex}}:\\ e\not \sim e'}} w_e w_{e'} \{P_e,P_{e'}\}.
\end{aligned}
\eeq
Next, we sum this identity over plaquettes $\hex\in \curly{P}_{m_1,m_2}$,
\beq\label{eq:AUB}
\curly{A}=\sum_{\minihex\in \curly{P}_{m_1,m_2}}\l(\tilde H_{\curly{F}_{\minihex}}+Q_{\curly{F}_{\minihex}}+R_{\curly{F}_{\minihex}}\r).
\eeq

We consider the sums over $\tilde H_{\curly{F}_{\minihex}}$, $Q_{\curly{F}_{\minihex}}$, and $R_{\curly{F}_{\minihex}}$ separately.\\

The sum over $\tilde H_{\curly{F}_{\minihex}}$ can be computed in the same way as the sum in \eqref{eq:sum}, with the only difference being that the weight $a$ is replaced by the weight $a^2$. This gives
\beq\label{eq:Hidentity}
\sum_{\minihex\in \curly{P}_{m_1,m_2}} \tilde H_{\curly{F}_{\minihex}}=(10+4a^2) H.
\eeq

We come to $\sum_{\minihex\in \curly{P}_{m_1,m_2}}Q_{\curly{F}_{\minihex}}$ next. This can be treated by similar considerations, except that we are now counting pairs of distinct edges $e\sim e'$. From Definition \eqref{eq:QRminihexdefn} and translation invariance, we see that this gives a multiple of $Q$ defined in \eqref{eq:QRdefn}. To find the combinatorial prefactor, we count how often a pair of edges $e\sim e'$ (i.e., a pair of distinct edges sharing a single vertex) appears in a copy of $\curly{F}_{\minihex}$, taking into account the weight factor $w_e w_{e'}$ as well. We find that each pair of edges $e\sim e'$ appears in $7$ copies of $\curly{F}_{\minihex}$ without weights, in $2$ copies with one of the edges weighted, and in $3$ copies with both edges weighted. This implies that
\beq\label{eq:Qidentity}
\sum_{\minihex\in \curly{P}_{m_1,m_2}}
Q_{\curly{F}_{\minihex}}
=(3a^2+2a+7) Q.
\eeq

Finally, we consider the sum $\sum_{\minihex\in \curly{P}_{m_1,m_2}}R_{\curly{F}_{\minihex}}$. By \eqref{eq:QRminihexdefn}, the sum is over operators $\{P_e,P_{e'}\}$ for edges $e\not\sim e'$ not overlapping at a vertex. Hence, the two projections commute and we have
\beq\label{eq:commute}
\{P_e,P_{e'}\}=2P_e P_{e'}\geq 0.
\eeq
Next, we observe that the (weighted) number of times that a pair of edges $e\not\sim e'$ appears in a copy of $\curly{F}_{\minihex}$ is dominated by the number of times that a pair of edges $e\sim e'$ appears. Hence, the combinatorial considerations that led us to \eqref{eq:Qidentity} combined with \eqref{eq:commute} imply that
$$
\sum_{\minihex\in \curly{P}_{m_1,m_2}}
R_{\curly{F}_{\minihex}}
\leq (3a^2+2a+7) R.
$$
Returning to \eqref{eq:AUB} and applying this operator inequality as well as \eqref{eq:Qidentity}, we conclude that \eqref{eq:A2} holds. This proves Lemma \ref{lm:key}.
\e{proof}

\subsection{Concluding the finite-size criterion}

We are now ready to prove the finite-size criterion.

\be{proof}
We apply \eqref{eq:H2} followed by \eqref{eq:A2} and \eqref{eq:A1} to find
$$
\begin{aligned}
H^2
=&H+Q+R\\
\geq& H-\frac{10+4a^2}{3a^2+2a+7}H+\frac{1}{3a^2+2a+7}\curly{A},\\
\geq& H-\frac{10+4a^2}{3a^2+2a+7}H+\frac{10+4a}{3a^2+2a+7}\gam_{\curly{F}}H\\
=& \frac{10+4a}{3a^2+2a+7}\l(\gam_{\curly{F}}-\frac{a^2-2a+3}{10+4a}\r)H
\end{aligned}
$$
This proves \eqref{eq:H2claim} and hence the finite-size criterion.
\e{proof}

\subsection*{II. $\mathrm{SU}(2)$ symmetric DMRG for excited states}

\noindent
In terms of spin operators, the AKLT Hamiltonian is defined as~\cite{GMW}
\begin{eqnarray}
\nonumber
H_{\text{AKLT}}^{S=3/2}&=&\frac{27}{160}\sum_{\langle i,j\rangle}\big[
  \vec{S_i}\cdot\vec{S_j}+\frac{116}{243}(\vec{S_i}\cdot\vec{S_j})^2\\
&&+\frac{16}{243}(\vec{S_i}\cdot\vec{S_j})^3+\frac{55}{108}\big].
\label{aklthmlt}
\end{eqnarray}
When expanding in $S^{\pm}$ and $S^z$ operators, there will be $3+3^2+3^3$
distinguishable operator pairs per bond. Alternatively, one can formally sum
these operator pairs into a compact matrix product operator (MPO), paying the
price of generating a large MPO bond dimension $D=39$. Whichever option is
taken, the computation will be expensive. However, if the interaction is
written as a $\mathrm{SU}(2)$ invariant vector operator, the dimension is much smaller,
$D=11$, and the AKLT hamiltonian can be cast in a more convenient form and
treated much easier with the DMRG method. 

To facilitate understanding of how
this is done in practice, we present the following necessary but brief
introduction to the $\mathrm{SU}(2)$ invariant MPO and matrix product state (MPS) for
realizing the AKLT Hamiltonian. We do so without going too much into
algorithmic details that can be found in the literature
\cite{Weichselbaum2012} but focus on the steps directly related to our
implementation of the AKLT Hamiltonian and calculations of excited states by
exactly projecting out the massively degenerate ground state manifold in
systems with 'dangling' boundary spins, such as the 36-site cluster in
Fig.~2. We do not explain all terminology and presume that the reader has
sufficient familiarity with DMRG and MPS calculations.

An $\mathrm{SU}(2)$ invariant MPS can, loosely speaking, be made out of a summation of
different quantum fusion paths of a composite MPS constructed as a direct product 
of two layers of structureless (plain and without any symmetry) MPSs; a reduced layer 
and a Clebsch-Gordon coefficient (CGC) layer, the tensor product of which
guarantees a spin rotation invariant wavefunction. Locally, the $\mathrm{SU}(2)$
invariant $T_{i,j,k}$ tensor representing the local spin degrees of freedom is
also a summation of a direct product of the reduced plain tensor
$B(q_i,q_j,q_k)$ and the CGC tensor $C(z_j,z_j,z_k)$, with matching angular
momentum quantum numbers $q_i,q_j,q_k$ and their $z$ components
$z_i,z_j,z_k$. Quantum fluctuations allow various $q_i,q_k$ values to be
visited (assuming $q_j$ is local spin momentum that is fixed), corresponding to
all possible allowed fusion paths when forming a total angular momentum $J$ out
of the wavefunction.

\begin{figure}[t]
%\vspace{0.1cm}
\begin{center}
\includegraphics[width=75mm]{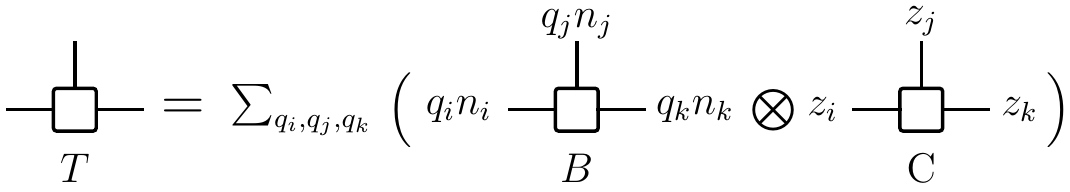}
\end{center}
\caption{Graphical representation of the local wavefunction $T$ of an $\mathrm{SU}(2)$ invariant MPS.}
\label{su2mps}
\end{figure}

\begin{figure}[b]
%\vspace{0.1cm}
\begin{center}
\includegraphics[width=75mm]{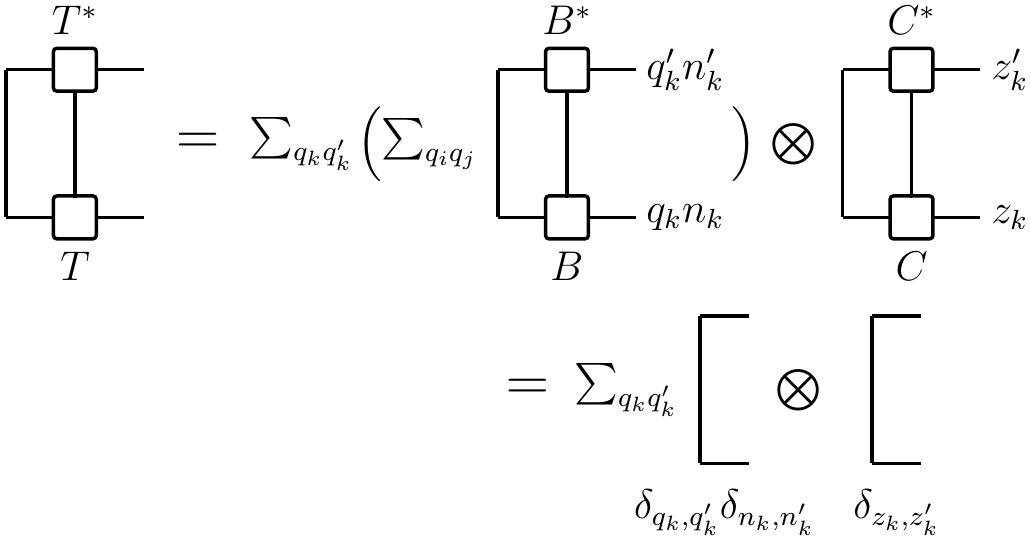}
\vskip-3mm
\end{center}
\caption{Left canonical constraint of the local wavefunction $T$ of an $\mathrm{SU}(2)$ invariant MPS.}
\label{su2mpscanonical}
\end{figure}

Given an angular moment fusion path $q_i\bigotimes q_j\to q_k$, the local reduced 
tensor $B(q_i,q_j,q_k)$ is written as
\begin{equation}
B(q_i,q_j,q_k)=\sum_{n_i,n_j,n_k}b^{q_kn_k}_{q_in_i,q_jn_j}|q_in_i,q_jn_j,q_kn_k\rangle,
\end{equation}
where $b^{q_kn_k}_{q_in_i,q_jn_j}$ is a coefficient, $n_i$ ($n_k$) is the
channel index which marks the path corresponding to the way in which spins to
the left (right) of the current one (along the 1D path of spins representing 
neighbors in the MPS) are fused into angular momentum $q_i$ ($q_k$). 
The subscripts $q_i,q_j$ indicate incoming angular momenta, and $q_k$ represents 
the outgoing angular momentum. The corresponding CGC matrices $C(z_i,z_j,z_k)$ are
\begin{equation}
C(z_i,z_j,z_k)=\sum_{z_i,z_j,z_k}c_{z_i,z_j}^{z_k}|z_i,z_j,z_k\rangle,
\end{equation}
where the CGCs $c_{z_i,z_j}^{z_k}$ satisfy the relation
\begin{equation}
\sum_{z_i,z_j}c^{z_k}_{z_i,z_j}c^{z_k^{\prime}}_{z_i,z_j}=\delta_{z_k,z^{{\prime}}_k}.
\end{equation}
Putting together $B$ and $C$, the local $\mathrm{SU}(2)$ invariant $T$ matrices of an MPS is
\begin{eqnarray}
\nonumber
T&=&\sum_{q_i,q_j,q_k}B(q_i,q_j,q_k)\bigotimes C(z_i,z_j,z_k)\\
\nonumber
&=&\sum_{q_i,q_j,q_k}\Big[\sum_{n_i,n_j,n_k}b^{q_kn_k}_{q_in_i,q_jn_j}|q_in_i,q_jn_j,q_kn_k\rangle\\
&&\bigotimes\sum_{z_i,z_j,z_k}c^{z_k}_{z_i,z_j}|z_i,z_j,z_k\rangle\Big],
\end{eqnarray}
whose graphical representation is shown in Fig.~\ref{su2mps}.

The splitting of a local $T$ matrix into a proper summation of the direct product of a reduced 
matrix and its CGC matrix greatly boosts the computational efficiency, reducing memory requests as 
well as ensuring the $\mathrm{SU}(2)$ spin rotation invariance.

The left canonical condition for an $\mathrm{SU}(2)$ symmetric MPS is depicted in Fig.~\ref{su2mpscanonical}. 
To arrive at the right hand side of this figure, the following left canonical constraint on the reduced matrices 
$B$ is imposed:
\begin{equation}
\sum_{q_i,q_j} \sum_{n_i,n_j} b^{*q_k^{\prime}n_k^{\prime}}_{q_in_i,q_jn_j} b^{q_kn_k}_{q_in_i,q_jn_j}=\delta_{q_k,q_k^{\prime}}\delta_{n_kn_k^{\prime}}.
\end{equation}
Similarly one can draw and write down the right canonical condition for $T$ matrices (omitted here).

\begin{figure}[t]
%\vspace{0.1cm}
\begin{center}
\includegraphics[width=75mm]{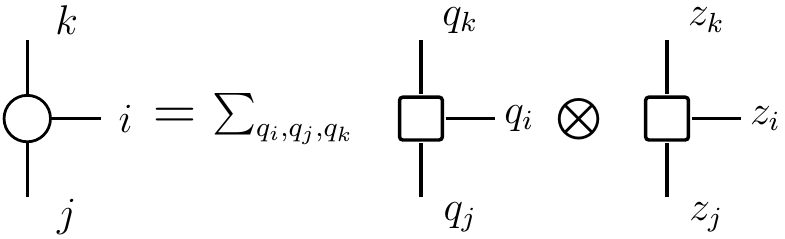}
\end{center}
\vskip-3mm
\caption{Graphical representation of a vector operator taking the form of a
  wavefunction, i.e., it can be decoupled into a direct product of a reduced
  operator matrix $\langle q_k\|\hat{O}_{q_i}\|q_j\rangle$ and its CGC.}
\label{su2bareoperator}
\end{figure}

Another important ingredient for realizing a $\mathrm{SU}(2)$ invariant MPS is the Wigner-Eckart theorem. 
It states that matrix element of a vector operator $\hat{O}$ which has angular moment $q_i$ and acting on a 
state with angular moment $q_j$ transforms under group generators like a wavefunction,
\begin{equation}
\langle q_kz_k|\hat{O}_{q_iz_i}|q_jz_j\rangle=\langle q_k\|\hat{O}_{q_i}\|q_j\rangle c_{z_i,z_j}^{z_k},
\end{equation}
where $c_{z_i,z_j}^{z_k}$ is the CGC and $\langle q_k\|\hat{O}_{q_i}\|q_j\rangle$ is a number that depends on
$\hat{O}_{q_i}$ and $q_j,q_k$. This condition means that one can write down a vector operator like a wavefunction, 
as in Fig.~\ref{su2bareoperator}.

\begin{figure}[b]
%\vspace{0.1cm}
\begin{center}
\includegraphics[width=80mm]{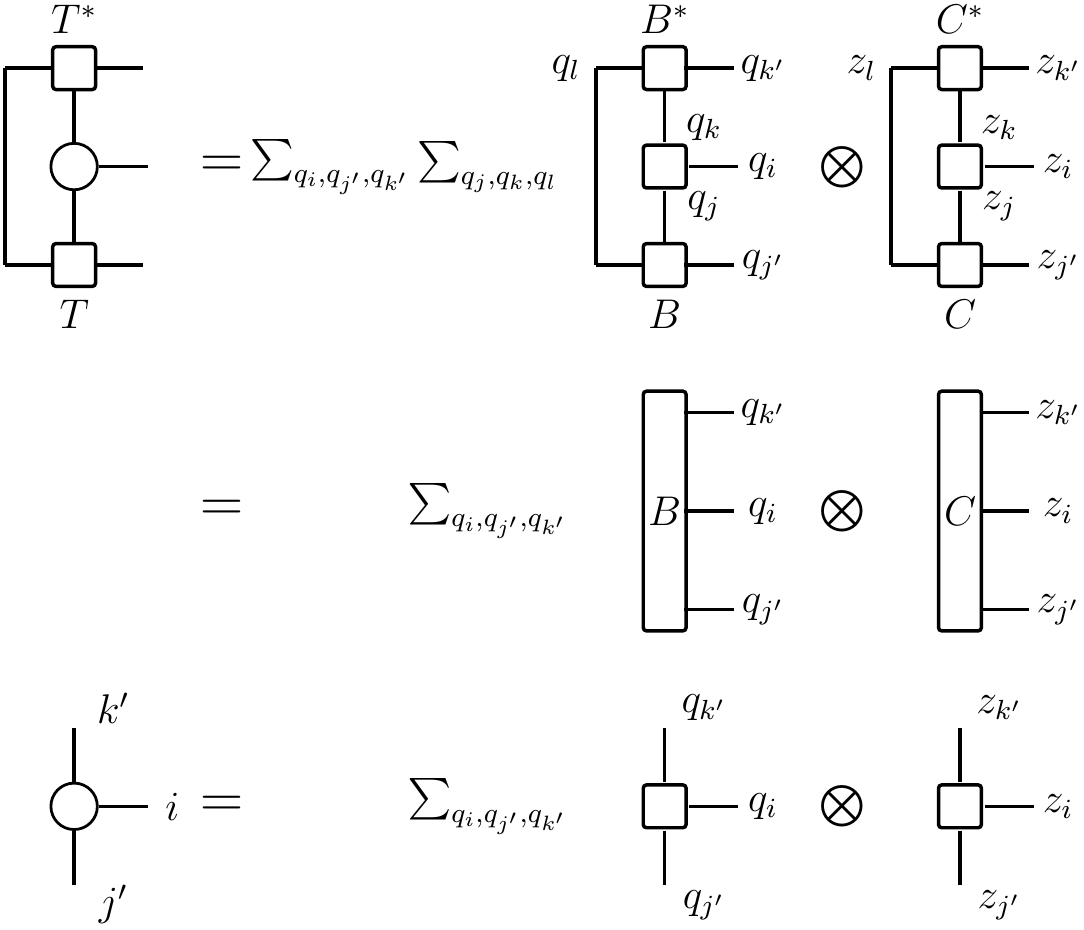}
\end{center}
\vskip-3mm
\caption{Illustration of a vector operator (as in Fig.~\ref{su2bareoperator}) under a basis
  transformation. It preserves the form of the $\mathrm{SU}(2)$ wavefunction.}
\label{su2operator}
\end{figure}

\begin{figure}[!t]
%\vspace{0.1cm}
\begin{center}
\includegraphics[width=80mm]{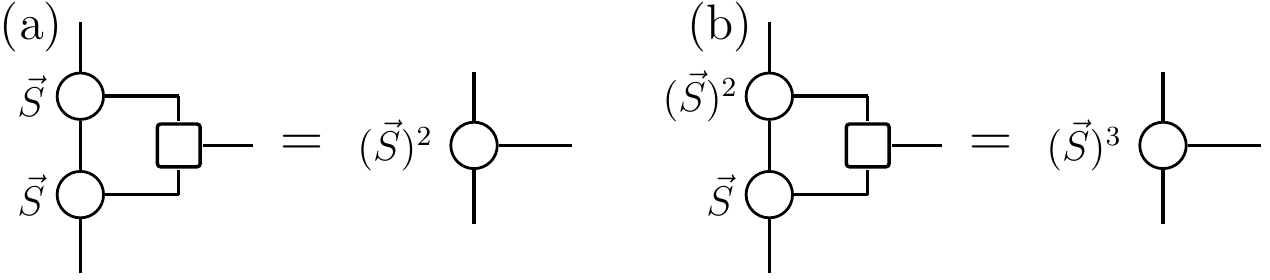}
\end{center}
\vskip-3mm
\caption{Demonstration on how to obtain $(\vec{S})^2$ and $(\vec{S})^3$
  operators via the fusion process. The fusion matrix (3-leg square tensor) is
  the identity matrix of the fusion process $q_i\bigotimes q_j=\bigoplus q_k$. }
\label{su2operatorfuse}
\end{figure}

\begin{figure*}[!t]
\begin{center}
\includegraphics[width=11cm]{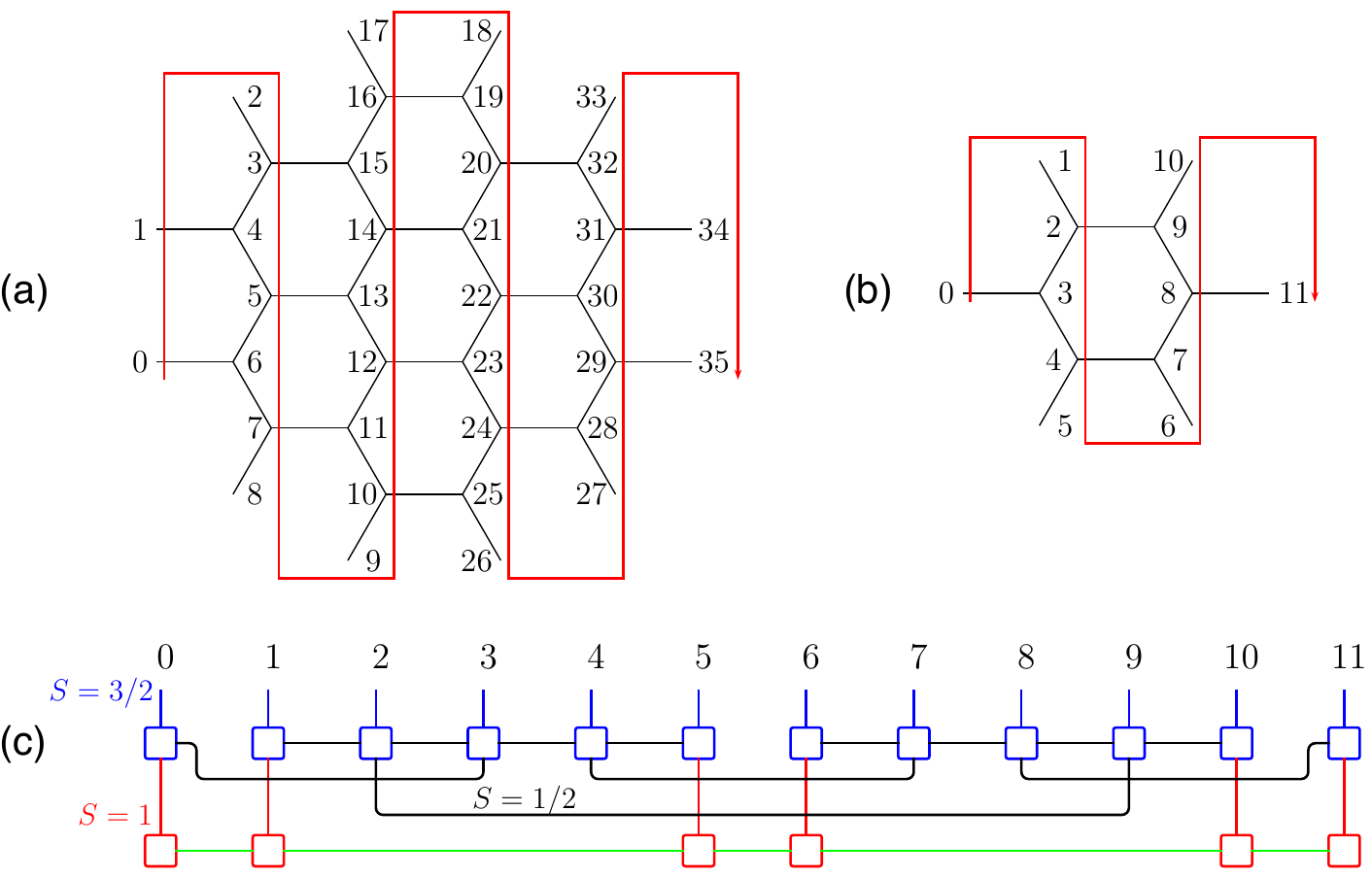}
\end{center}
\vskip-3mm
\caption{Path taken through a spin cluster in order to represent its ground state and excitations by MPSs;
(a) the 36-site cluster on which our proof is based and (b) a smaller  12-site illustrative cluster. The MPS representing 
a 2D AKLT state on the path is made out of two layers of MPSs in (c); the top layer (blue) reproduces the 2D lattice 
connectivity, with the boxes correspond to the tensors $T$ [which incorporate $\mathrm{SU}(2)$ symmetry via the $B$ and $C$
tensors discussed in the text]. In the lower layer, red lines and boxes represent $S=1$ free boundary spins and the green 
lines show one of the non-repeating paths that fuse all $S=1$ into a total angular moment $J$.}
\label{fig:snake}
\end{figure*}

\begin{figure}[b]
\begin{center}
\includegraphics[width=70mm]{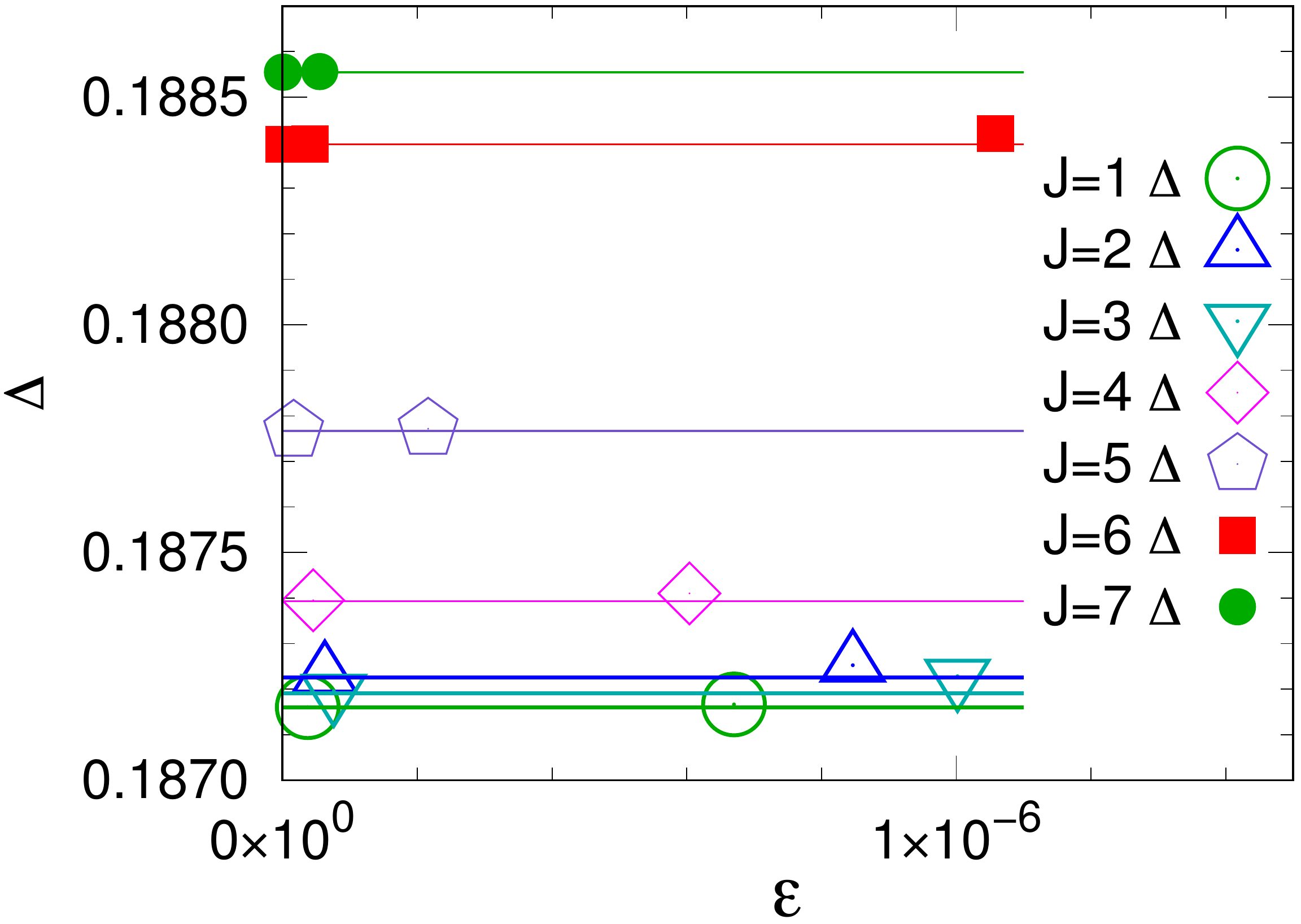}
\end{center}
\vskip-5mm
\caption{The smallest nonzero gaps in the sectors $1\le J\le L+1=7$ vs the discarded weight in DMRG 
calculations for the open-boundary 12-sites AKLT cluster in Fig.~\ref{fig:snake}(b) with the bond weights
in Eq.~(\ref{hakltedges}) taken to be $w_e=1.2$ on the central hexagon and $w_e=1$ on the edges connecting
to the boundary dangling spins. The solid lines indicate the corresponding exact results from Lanczos diagonalization. 
The case $J=0$ is not shown here because that gap is much larger, but our method also reproduces it very well. The
smallest gap is in the $J=1$ sector.}
\label{fig:dmrg12}
\end{figure}

\begin{figure*}[!t]
\begin{center}
\includegraphics[width=14cm]{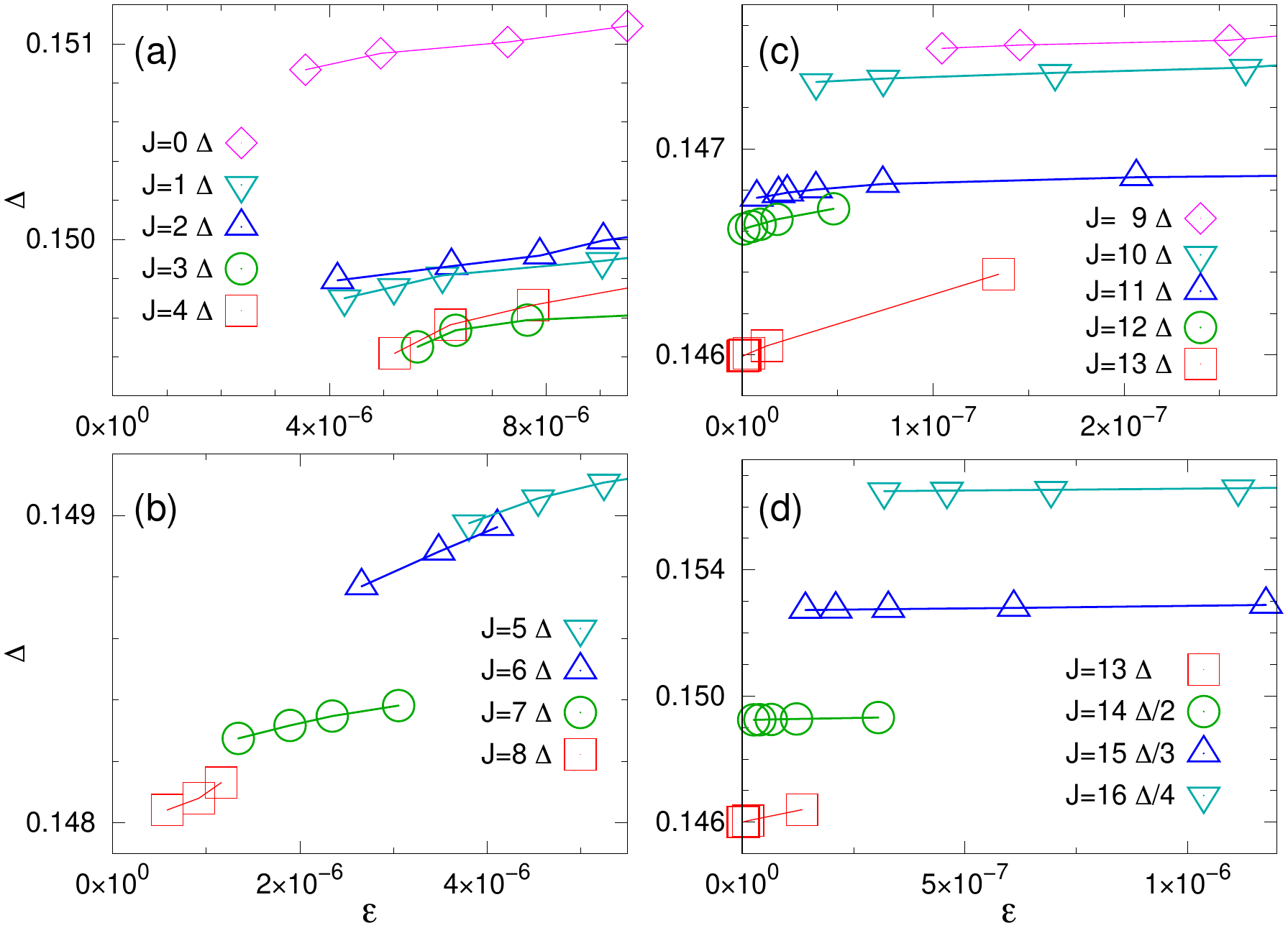}
\end{center}
\vskip-5mm
\caption{The smallest nonzero gaps in the sectors $0\le J\le L+4=16$ vs
  the discarded weight in DMRG calculations for the open-boundary 36-sites cluster
  depicted in Fig.~2 of the main paper. Here the value of the adjustable bulk coupling
  is $a=1.4$, which is the value for which we have proved the finite-size criterion. 
  The solid lines drawn through the $J=13$ points in (b) and (d) are linear fits giving
  the extrapolated gap $\Delta(13)=0.14599$. The lines between points for other $J$
  values are only guides to the eye. Note that, in (d) the gaps have been divided by
  $J-12$ in order to compress the horizontal scale (also demonstrating that the
  gaps for large $J$ scale roughly as $J+1-L$ in this case). The maximum bond dimension (corresponding
  to the smallest $\epsilon$ for each $J$) is $D=2000$ in panel (a), $D=2400$ for $5\leq J\leq 12$, and $D=1200$ for all other
  cases.}
\label{fig:dmrg}
\end{figure*}

Given an operator $\hat{O}_{q_i}$ in its $\mathrm{SU}(2)$ invariant form, its basis transformation is 
guaranteed to preserve the same form, as demonstrated in Fig.~\ref{su2operator}.
The spin operator 
\begin{equation}
\left(-\frac{1}{\sqrt{2}}S^+,S^z,\frac{1}{\sqrt{2}}S^-\right)^{T}
\end{equation}
transforms like a wavefunction with angular moment 1. To realize the AKLT
Hamiltonian Eq.~\eqref{aklthmlt} written in terms of vector operators $(\vec{S})^1$,
$(\vec{S})^2$, and $(\vec{S})^3$, requires implementing their $\mathrm{SU}(2)$
invarient representations. For example, $(\vec{S}^2)$ is constructed simply by
multiplying two $\vec{S}$ operators on the physical index and successively
fusing two $q=1$ angular momenta on the virtual index into a total angular
momentum $q=0,1,2$, following the fusion rule $1\otimes 1=0\oplus 1\oplus 2$,
as shown in Fig.~\ref{su2operatorfuse}(a). Similarly, $(\vec{S})^3$ can be
constructed by multiplying $(\vec{S})^2$ and $\vec{S}$ on the physical index
and fusing the virtual index, as in Fig.~\ref{su2operatorfuse}(b).

With the above preparation in an $\mathrm{SU}(2)$ invariant basis, one can enumerate the complete 
ground state manifold and computate the lowest excitation in each total spin $J$ sectors, 
which we here do for the 36-site 2D AKLT cluster depicted in Fig.~2 in the main text. 
For illustration purposes we will here also consider a 12-site cluster, for which it is 
easier to draw pictures of the MPSs incorporating the edge spins; Fig.~\ref{fig:snake}.

The degenerate ground states of the cluster with open boundaries are generated by 
first preparing them in their 2D tensor network representation, as with the black solid lines 
in Figs.~\ref{fig:snake}(a) and (b). Then, as always in 2D DMRG calculations, a path is
chosen to 'snake' through the 2D network to compress the states into MPSs. The paths 
chosen here for the two clusters are indicated with red lines. This type of path represents
the minimum number of cuts when partitioning the system into two arbitrary parts. Minimizing 
the number of cuts optimizes the ability of the MPS to build in bipartite entanglement. 

The compression procedure 
and special treatment of the boundary spins works as follows, using the 12-site system for 
definiteness in the illustration in Fig.~\ref{fig:snake}(c). First, the 'snake' is stretched 
into a line as shown with the blue boxes, which represent the $\mathrm{SU}(2)$ $T$ tensors discussed 
above. The connectivity of the original 2D network is shown with the black lines. The 
remaining dangling $S=1$ degrees of freedom form a set of unitary orthogonal MPSs with 
different total angular momenta; this MPS representation is shown with the red boxes. The
green lines can connect these boxes in any non-repeating order, and a given path corresponds
to a set of fusion values that define the quantum numbers associated with the line segments.
The final MPS representing the 2D AKLT ground state is formulated by combining the two 
layers of matrix product states as in Fig.~\ref{fig:snake}(c); a blue layer of all physical 
spins and a red layer of the dangling boundary spins. 

All the paths [green lines in Fig.~\ref{fig:snake}(c)] connecting 
the tensors of the lower layer have to be considered to construct the full ground state manifold.
Generating these unitary orthogonal MPSs of the free $S=1$ boundary degrees of freedom is a 
computer facilitated automatic process that requires a computational effort scaling with
the Hilbert space sizes for all possible $J$---these sizes are listed for the 36-site cluster in the main paper. 
Once all the ground state in a given sector $J$ has been gathered, one can employ the DMRG 
algorithm for excited states, as described in Ref.~\cite{prl121.107202}, to compute the
first excited state above the ground state manifold. In the case considered here the procedure 
is simplified due to the fact that the ground states are known exactly and are written out 
straight forwardly without any energy minimization.

\subsection*{III. DMRG gap convergence }

\noindent
We have carried out various tests to confirm the correctness of the DMRG code, e.g., 
using the 1D AKLT chain and smaller 2D clusters for which the gaps can be verified using Lanczos 
exact diagonalization. Even with the $\mathrm{SU}(2)$ symmetry implemented and all the degenerate ground state 
projected out exactly, reliably computing the gaps for all $J$ values of interest for a cluster
with $36$ spins is not an easy task. Convergence as a function of the bond dimension $D$ has to be 
carefully checked. Instead of monitoring the convergence directly versus $D$, it is better to consider 
the energy as a function of the discarded weight $\epsilon$ of the DMRG procedure obtained for each 
$D$ used. 

For the 12-site cluster in Fig.~\ref{fig:snake} we can easily compute the ground state
in each $J$ sector by Lanczos exact diagonalization. We can then unambiguously test our DMRG method
with the MPS-expressed degenerate ground states projected out exactly. Fig.~\ref{fig:dmrg12}
shows the results for several $J$ values versus the discarded DMRG weight $\epsilon$. The solid lines
are the exact Lanczos results, and they agree very well with the DMRG results for small $\epsilon$.
For this cluster the lowest gap is in the $J=1$ sector and there is no particular systematic ordering
of the levels.

In Fig.~\ref{fig:dmrg} we show our results for the 36-site cluster. As discussed in the main 
text, we expect the smallest gap should be for $J \in \{0,1,\ldots,13\}$, but we carried out 
calculations for all possible $J$-values and confirmed that the gaps increase rapidly upon increasing $J$ 
above $J=13$. Since the values of $\epsilon$ for which extrapolations can reliably be carried out 
span a wide range, rather systematically depending on $J$, in Fig.~\ref{fig:dmrg}  we have divided up the 
results for the different $J$ values into four different panels with groups of similar $J$ values. The 
reason for the larger $\epsilon$ for smaller $J$ is primarily due to the size of the Hilbert space,
which increases with decreasing $J$.

Based on these results there is no doubt that the smallest gap of this cluster 
is in the $J=13$ sector. The gaps increase rapidly with $J$.  We mention that the gaps in sectors of very large $J$ can also be estimated analytically, e.g., by using the projection lemma from \cite{KKR}. For smaller $J$ 
the gaps initially increase monotonically, but for the $J \le 4$ non-monotonic behavior 
sets in. There the gaps are already much larger than $\Delta(13)$.

\end{document}